\documentclass[prd,twocolumn,showpacs,preprintnumbers,amsmath,nofootinbib,amssymb,floatfix]{revtex4-1}
\usepackage{graphicx}
\usepackage[sort&compress]{natbib}
\usepackage{subfigure}
\usepackage{amsmath}
\usepackage{amsfonts}
\usepackage{cancel}
\usepackage{lmodern,dsfont}
\usepackage{amssymb}
\begin{document}
\arraycolsep1.5pt
\newcommand{\Ima}{\textrm{Im}}
\newcommand{\Rea}{\textrm{Re}}
\newcommand{\mev}{\textrm{ MeV}}
\newcommand{\be}{\begin{equation}}
\newcommand{\ee}{\end{equation}}
\newcommand{\ba}{\begin{eqnarray}}
\newcommand{\ea}{\end{eqnarray}}
\newcommand{\gev}{\textrm{ GeV}}
\newcommand{\nn}{{\nonumber}}
\newcommand{\dtres}{d^{\hspace{0.1mm} 3}\hspace{-0.5mm}}
\newcommand{\rts}{ \sqrt s}
\newcommand{\non}{\nonumber \\[2mm]}

\title{
Finite volume treatment of $\pi\pi$ scattering and limits to phase
shifts extraction from lattice QCD}

\author{M.~Albaladejo$^1$, J.~A.~Oller$^1$, E. Oset$^2$, G.~Rios$^1$ and L. Roca$^1$
}
\affiliation{
$^1$Departamento de F\'{\i}sica. Universidad de Murcia. E-30100 Murcia, Spain.\\
$^2$Departamento de F\'{\i}sica Te\'orica and IFIC, Centro Mixto Universidad de Valencia-CSIC,
Institutos de Investigaci\'on de Paterna, Aptdo. 22085, 46071 Valencia,
Spain.
}

\date{\today}

\begin{abstract}

  We study theoretically the effects of finite volume for   $\pi\pi$
scattering in order to extract physical observables for infinite volume
from lattice QCD. We compare three different approaches for $\pi\pi$
scattering (lowest order Bethe-Salpeter approach, $N/D$ and
inverse amplitude methods) with the aim to study the effects 
of the finite
size of the box in the potential of the different theories, specially
the left-hand cut contribution 
through loops in the crossed $t,u-$channels.
 We quantify the error made by
neglecting these effects in usual
extractions of 
physical observables from lattice QCD spectra.  We conclude that for $\pi\pi$ phase-shifts in the
scalar-isoscalar channel up to $800\mev$ this effect is negligible for
box sizes bigger than $2.5m_\pi^{-1}$ and of the order of 5\% at around
$1.5-2m_\pi^{-1}$. For isospin 2 the finite size effects can reach up to
10\% for that energy. We also quantify the error made when using the standard  
L\"uscher method to extract physical observables from lattice QCD, 
which is widely used in the literature but is an approximation of the one used in the
present work.

\end{abstract}

\maketitle

\section{Introduction}
\label{Intro}

One of the aims in present lattice QCD calculations is the determination
of the hadron spectrum and many efforts are devoted to this task
\cite{Nakahara:1999vy,Mathur:2006bs,Basak:2007kj,Bulava:2010yg,Morningstar:2010ae,Foley:2010te,Alford:2000mm,Kunihiro:2003yj,Suganuma:2005ds,Hart:2006ps,Wada:2007cp,Prelovsek:2010gm,Lin:2008pr,Gattringer:2008vj,Engel:2010my,Mahbub:2010me,Edwards:2011jj,Lang:2011mn,Prelovsek:2011im}.
A recent review on the different methods used and results can be seen in
\cite{reviewhadron}. Since one evaluates the spectrum 
for particles in a
finite box, one must use a link from this spectrum to the physical one
in infinite space. Sometimes, when it rarely happens, an energy of
the box rather independent of the volume is taken as a proof that this
is the energy of a state in the infinite volume space.   In other works
the  ``avoided level crossing'', with lines of spectra that get close to
each other and then separate, is usually taken as a signal of a
resonance, but this criteria has been shown insufficient for resonances
with a large width \cite{Bernard:2007cm,Bernard:2008ax,misha}. A more
accurate method consists on the use of L\"uscher's approach, but this
works for resonances with only one decay channel. The method allows to
reproduce the phase-shifts for the particles of this decay channel
starting from the discrete energy levels in the box
\cite{luscher,Luscher:1990ux}. This method has been recently simplified
and improved in \cite{misha} by keeping the full relativistic two-body
propagator (L\"uscher's approach keeps the imaginary part of this
propagator exactly but makes approximations on the real part). The work
of \cite{misha} also extends the method to two or more coupled channels.
The extension to coupled channels has also been worked out in
\cite{Liu:2005kr,Lage:2009zv,Bernard:2010fp}. The work of \cite{misha}
presents an independent method, which is rather practical, and has been
tested and proved to work in realistic cases of likely lattice results.
The method has been extended in \cite{mishajuelich} to obtain finite
volume results from the J\"ulich model for the meson-baryon interaction
and in \cite{alberto} to study the interaction of the $DK$ and $\eta
D_s$ system where the $D_{s^*0}(2317)$ resonance is dynamically
generated from the interaction of these particles. The case of the
$\kappa$ resonance in the $K \pi$ channel is also addressed  in \cite{mishakappa} following 
the approach of Ref.~\cite{misha}. It has also been extended to
the case of interaction of unstable particles in \cite{luis}, to the
study of the DN interaction \cite{xie}, the $\pi \pi$ interaction in the
$\rho$ channel \cite{chen} and to find strategies to determine the two
$\Lambda(1405)$ states from lattice results \cite{lambda}.

In Ref.~\cite{misha} the
 problem of getting phase-shifts and resonances 
from lattice QCD results (``inverse problem'') using two coupled channels was addressed. Special attention was given to the evaluation of errors and the precision needed
on the lattice QCD calculations to obtain phase-shifts and resonance
properties with a desired accuracy. The derivation of the basic formula
of \cite{misha} is done using the method of the chiral unitary approach
\cite{review} to obtain the scattering matrix from a potential. This
method uses a dispersion relation for the inverse of the amplitude,
taking the imaginary part of $T^{-1}$ in the physical region and using
unitarity in coupled channels \cite{nsd,ollerulf}. The method does not
integrate explicitly over the left-hand cut singularity. Nevertheless, 
the latter might lead to interesting problems in finite volume calculations because in field
theory, loops in the $t-$ or $u-$channel that contribute to crossed cuts,
are volume dependent. There is no problem to incorporate these extra
terms into the chiral unitary approach by putting them properly in the 
interaction kernel of the Bethe Salpeter equation or N/D method
\cite{nsd,guo},  or using the inverse amplitude method (IAM)
\cite{Dobado:1996ps,ramonetiam}. However,  the method of \cite{misha} to
analyze lattice spectra and obtain phase-shifts explicitly relies upon
having a kernel in the Bethe Salpeter equation which is volume
independent. The same handicap occurs in the use of the standard
L\"uscher approach, where contributions from possible volume dependence
in the potential are shown to be ``exponentially suppressed''
 in the box
volume. Yet, there is no way, unless one knows precisely the source of
the volume dependent terms, to estimate these effects and determine for
which volumes the ``exponentially suppressed'' corrections have become
smaller than a desired quantity. This is however an important
information in realistic calculations. The purpose of the present paper
is to address this problem in a practical case, the scattering of pions
in s-wave. For that we determine the strength of these  volume dependent
terms as a function of the size of the box and the impact of these
effects in the determination of the phase-shifts in the infinite volume
case. 

The contents of the paper are as follows. After this introduction, we summarize 
in Sec.~\ref{sec2} the three models used to evaluate $\pi\pi$ scattering in the 
infinite and finite volume case. We then follow by studying the dependence on the lattice 
size of the box $L$ of the resulting phase shifts in Sec.~\ref{sec3}. Conclusions are collected in 
Sec.~\ref{sec4}.

%%%%%%%%%%%%%%%%%%%%%%%%%%%%%%%%%%%%%%%%%%%%%%%%%%%%%%%%%%%%%%%%%%%%%%%%%%%%%%%
%%%%%%%%%%%%%%%%%%%%%%%%%%%%%%%%%%%%%%%%%%%%%%%%%%%%%%%%%%%%%%%%%%%%%%%%%%%%%%%
\section{The $\pi\pi$ scattering in the finite box}
\label{sec2}

In this section we explain the three models that 
we are going to consider in the
present work to evaluate the $\pi\pi$ scattering within the chiral
unitary approach: lowest order Bethe-Salpeter (BS), N/D and Inverse 
amplitude method (IAM). The latter two provide contributions
to the left-hand cut of the scattering amplitude while the BS does not.
After summarizing the models for the 
infinite  volume, we explain for each of them how to evaluate the
scattering in a box of finite size $L$. We  study the scalar
channel up to total energies of about $800\mev$ for both isospin ($I$) 0 and 2. 
 The isoscalar case is relevant for the lattice QCD studies of
$\sigma$ (or $f_0(600)$ \cite{pdg})  meson resonance, while for the isotensor case 
 the left-hand cut is more relevant (see below).
Up to those energies the $K\bar K$ and $\eta\eta$ channels in the $I=0$
case are negligible, hence, we deal here only with the $\pi\pi$ channel.

%%%%%%%%%%%%%%%%%%%%%%%%%%%%%%%%%%%%%%%%%%%%%%%%%%%%%%%%%%%%%%%%%%%%%%%%%%%%%%%%%%%%%
\subsection{Lowest order Bethe-Salpeter approach}
\label{sec:BS}

In the chiral unitary approach the scattering matrix 
can be  given by the Bethe-Salpeter equation in its factorized form
\cite{npa}

\be
T=[1-VG]^{-1}V= [V^{-1}-G]^{-1}~,
\label{bse}
\ee
where $V$ is the $\pi\pi$ potential,
$V=-\frac{1}{f_\pi^2}(s-\frac{m^2}{2})$ for $I=0$ and 
$V=\frac{1}{2f_\pi^2}(s-2 m^2)$ for $I=2$,
which are obtained from the lowest order
chiral Lagrangians \cite{Gasser:1983yg},
with $m$ the pion mass and $f_\pi=92.4\mev$.
In Eq.~(\ref{bse})
$G$ is the loop function of two meson propagators,
which is defined as 
\be
\label{loop}
G=i\,\int\frac{d^4 p}{(2\pi)^4} \,
\frac{1}{(P-p)^2-m^2+i\epsilon}\,\frac{1}{p^2-m^2+i\epsilon}
\ ,
%\label{eq:Gl}
\ee

with  $P$ the four-momentum of the global meson-meson system.
Note that Eq.~(\ref{bse}) only has right-hand
 cut, unlike the
other two approaches discussed in the next subsections.

The loop function in Eq.~(\ref{loop}) can be regularized
either with dimensional regularization
or with a three-momentum
cutoff. The connection between  both methods was
shown in Refs.~\cite{ollerulf,ramonetiam}.
In dimensional regularization\footnote{In our context we refer to the $G$ function given in Eq.~\eqref{eq:g-function} as calculated 
in ``dimensional regularization''. Of course, with the latter procedure the results is infinite. The infinite is removed by the subtraction constant $a(\mu)$. A more accurate formulation can be given in terms of dispersion relations, the interested reader on this point can consult Refs.~\cite{nsd,ollerulf}, though the final result is the same.} the integral of Eq.~(\ref{loop}), $G^D$, is evaluated and gives for the $\pi\pi$ system
\cite{ollerulf,bennhold}

\be
G^{\rm{D}}(E) =  \frac{1}{(4 \pi)^2}
 \Biggr\{
        a(\mu) + \log \frac{m^2}{\mu^2}+\sigma
	\log \frac{\sigma+1}{\sigma-1}
  \Biggr\}~,
\label{eq:g-function}
\ee
where $\sigma=\sqrt{1-\frac{4m^2}{s}}$, $s=E^2$, with $E$ the energy of the system in the center of mass
frame, $\mu$ is  a renormalization scale and $a(\mu)$ is a subtraction constant
(note that only the combination $a(\mu)-\log \mu^2$ is the relevant degree of freedom,  that is, there is only one independent
parameter).

The loop function $G$ can also be regularized with a three momentum
cutoff $p_{\rm max}$ and,
after the $p^0$ integration is performed \cite{npa}, it results

\ba
&&G(s)=\hspace{-4mm}\int\limits_{|\vec p|<p_{\rm max}}
\frac{d^3\vec p}{(2\pi)^3}\frac{1}{\omega(\vec p)}
\frac{1}
{s-4\omega(\vec p)^2+i\epsilon}~,
\non 
&&\omega(\vec p)=\sqrt{m^2+\vec p^{\,\,2}}~ .
\label{prop_cont}
\ea

Let us now address the modifications in order to evaluate the $\pi\pi$
scattering in a finite box following 
the procedure explained in Ref.~\cite{misha}.
The main difference with respect to the infinite volume case is that
 instead of integrating over the
energy states of the continuum with $\vec{p}$ being a continuous variable
as in Eq.~(\ref{prop_cont}), one must sum over 
the discrete momenta allowed
in a finite box of side $L$ with periodic boundary conditions.
We then have to replace $G$ by 
$\widetilde G$, where 
\ba
\widetilde G&=&\frac{1}{L^3}\sum_{\vec p}^{|\vec p|<p_{\rm max}}
\frac{1}{\omega(\vec p)}\,\,
\frac{1}
{s-4\omega(\vec p)^2}~,
\non 
\vec p&=&\frac{2\pi}{L}\,\vec n~,
\quad\vec n\in \mathds{Z}^3 \,
\label{tildeg}
\ea
For the sake of comparison with the other models considered in the
present work, where dimensional regularization is always done, we use
the procedure of \cite{alberto} in order to write the finite volume loop
function $\widetilde{G}$ in terms of the infinite volume one $G^D$ evaluated in dimensional
regularization:
\begin{align}
\widetilde G=G^{\rm{D}}+\lim_{p_\textrm{max}\to \infty}
\Bigg[\frac{1}{L^3}\sum_{p_i}^{p_\textrm{max}}I(p_i,s)
-\int\limits_{p<p_\textrm{max}}\frac{d^3p}{(2\pi)^3} I(p,s)\Bigg]~,
\label{tonediff}
\end{align}
where $I(p,s)$ is the integrand of Eq.~(\ref{prop_cont})

\be
I(p,s)=
\frac{1}{\omega(\vec p)}\,\,
\frac{1}
{s-4\omega(\vec p)^2}~.
\label{prop_contado}
\ee

Note that $\widetilde G$ of Eq.~(\ref{tonediff})
 depends on the subtraction constant $a$ instead of the
three-momentum cutoff $p_\textrm{max}$.

In the box the scattering matrix reads
\be
\widetilde T=\frac{1}{V^{-1}-\widetilde G}~.
\ee

The eigenenergies of
the box correspond to energies  that produce poles in the 
$\widetilde T$ matrix, 
which corresponds to the condition
$\widetilde G(E)=V^{-1}(E)$.
Therefore for those values of the energies, 
the $T$ matrix for infinite volume can be obtained
by 

   \be
T(E)=\left(V^{-1}(E)-G(E)\right)^{-1}= 
\left(\widetilde G(E)-G(E)\right)^{-1} \ . 
\label{extracted_1_channel}
\ee 
The amplitude is related to the phase-shifts by
\ba
T(E)=-\frac{8\pi E}{p}\frac{1}{\cot \delta-i}~,
\label{eq:defas}
\ea
where $p=\frac{E}{2}\sqrt{1-\frac{4m^2}{s}}$ 
is the CM momentum.

Eq.~(\ref{extracted_1_channel}) is nothing but L\"uscher formula
\cite{luscher,Luscher:1990ux} except that, as shown in
Ref.~\cite{misha}, Eq.~(\ref{extracted_1_channel}) 
keeps all the terms of the relativistic two-body propagator, while  
L\"uscher's approach neglects terms in $\textrm{Re}~I(p)$ which are
exponentially suppressed in the physical region, but can become sizable
below threshold, or in other cases when small volumes are used or large
energies are involved.

\subsection{The IAM approach}

The next approach considered is the elastic 
Inverse Amplitude Method (IAM)~\cite{Dobado:1996ps}, 
which we briefly review in this section and describe
how to extend it to consider scattering in a finite box. 

The elastic IAM makes use of elastic unitarity
and Chiral Perturbation Theory (ChPT)~\cite{Gasser:1983yg} 
to evaluate a dispersion relation for
the inverse of the $\pi\pi$ scattering partial wave of definite isospin $I$ and angular
momentum $J$, $T^{IJ}$ (in the following we drop the superscript 
$IJ$ to simplify  notation). The advantage of using the inverse of a partial wave
stems from the fact that its imaginary part is fixed by unitarity,
\begin{equation}
  \label{eq:unitarity}
  \Ima\,T=-\frac{\sigma}{16\pi}\vert T\vert^2\quad\Rightarrow\quad 
  \Ima\,T^{-1}=\frac{\sigma}{16\pi}~.
\end{equation}
Thus, the right-hand cut integral can be evaluated exactly in the elastic regime
and the obtained partial wave satisfies unitarity exactly.
 The partial wave amplitudes calculated in ChPT  cannot satisfy unitarity exactly
since they are obtained in a perturbative expansion $T= T_2+T_4+{\cal O}(p^6)$, 
where $T_2={\cal O}(p^2)$ and $T_4={\cal O}(p^4)$ are the Leading Order and
Next--to--Leading Order contributions in the chiral expansion of $T$, respectively.
However, unitarity is satisfied in a perturbative way,
\begin{equation}
  \label{eq:unitarity-pert}
\Ima \,T_2=0,\quad\Ima \,T_4=-\frac{\sigma}{16\pi} T_2^2,\quad\cdots.
\end{equation}
These equations allow us to evaluate the dispersion relation and obtain a
compact form for the partial wave as we show below.

We write then a dispersion relation for an auxiliary function $F\equiv T_2^2/T$,
whose analytic structure consists on a right-hand cut ($RC$) from $4m_\pi^2$ to $\infty$,
a left-hand cut ($LC$) from $-\infty$ to $0$, and possible poles coming
from zeros of $T$,
\begin{equation}
  \begin{aligned}
  \label{eq:disp-rel}
  F(s)&=F(0)+F'(0)s+\tfrac1{2}F''(0)s^2\\
  &+\frac{s^3}{\pi}\int_{RC}ds'\frac{\Ima\, F(s')}{s'^3(s'-s)}+LC(F)+PC~,
  \end{aligned}
\end{equation}
where we have performed three subtractions to ensure convergence.
In the above equation $LC(F)$ stands for the integral over the left-hand cut,
and $PC$ stands for possible poles contributions, which 
are present in the scalar waves due to the Adler zeros.
Using Eqs. \eqref{eq:unitarity} and \eqref{eq:unitarity-pert}
we can evaluate \emph{exactly} in the $RC$ integral $\Ima\,F=-\Ima\,T_4$,
and obtain for the right-hand cut $RC(F)=-RC(T_4)$.
The subtraction constants can be evaluated with ChPT since they only
involve amplitudes or their derivatives evaluated at $s=0$,
$F(0)\simeq T_2(0)-T_4(0)$, $F'(0)\simeq T_2'(0)-T_4'(0)$,
$F''(0)\simeq -T_4''(0)$. The left-hand cut can be considered to be 
dominated by its low energy part, since we have three subtractions,
and it is also  dumped by an extra $1/(s'-s)$ when considering physical values of $s$.
Then, we evaluate it using ChPT to obtain $LC(F)\simeq-LC(T_4)$.
The pole contribution is formally ${\cal O}(p^6)$ and we neglect it 
(this causes some technical problems in the subthreshold region around the
Adler zeros which can be easily solved, but they do not
affect the description of scattering or resonances, for details see~\cite{mIAM}).
Taking into account all the above considerations we arrive at
\begin{equation}
  \begin{aligned}
    \label{eq:preIAM}
    \frac{T_2^2(s)}{T(s)}&\simeq T_2(0)+T_2'(0)s-T_4(0)-T_4'(0)s-\tfrac1{2}T_4''(0)s^2\\
    &-RC(T_4)-LC(T_4)=T_2(s)-T_4(s)~,
  \end{aligned}
\end{equation}
where in the last step 
we have taken into account that $T_2(s)$ is just a first order polynomial
in $s$ so that $T_2(s)=T_2(0)+T_2'(0)s$, and that the remaining piece in the
middle member of Eq.~\eqref{eq:preIAM} is a dispersion relation
for $-T_4(s)$. Then one obtains the simple IAM formula, 
\begin{equation}
  \label{eq:IAM}
  T^{IAM}=\frac{T_2^2}{T_2-T_4}~.
\end{equation}
This formula can be systematically extended to higher orders by 
evaluating the subtraction constants and the left-hand cut in the dispersion
relation to higher orders. Note that the full one--loop ChPT calculation is used,
so the IAM partial waves depend on the chiral Low Energy Constants (LECs),
that absorb the loop divergences through their renormalization and
depend on a renormalization scale $\mu$. Of course, this $\mu$ dependence
is canceled out in physical observables. 
In the case of $\pi\pi$
scattering there appear four LECs, denoted $l_i^r(\mu)$, $i=1...4$.
These LECs are not fixed from symmetry considerations and their value
has to be determined from experiment. For the IAM calculations here we take the values used in \cite{Hanhart:2008mx}:
$10^3l_1^r=-3.7\pm  0.2$, $10^3l_2^r=5.0\pm 0.4$, $10^3l_3^r=0.8\pm 3.8$,
$10^3l_4^r=6.2\pm 5.7$, at $\mu=770$ MeV, which give a good description of
phase-shift
data. Note that in the present work
we are not interested in a detailed description of
scattering data, but on the effects of ignoring the exponentially suppressed dependence
on the box size when using L\"uscher's or the chiral unitary approach
to obtain the scattering phase-shifts from the energy levels in finite volume. 

To evaluate the IAM partial waves in a finite box of size $L$ we proceed
in a similar way as in 
subsection~\ref{sec:BS}.
The only  
piece we need to change is $T_4$, that receives 
contributions from loop diagrams, whose momentum integrals 
should be replaced by discrete sums over the
allowed momenta in the finite box. 

\begin{figure}[t]
  \centering
  \includegraphics[width=0.45\textwidth,keepaspectratio]{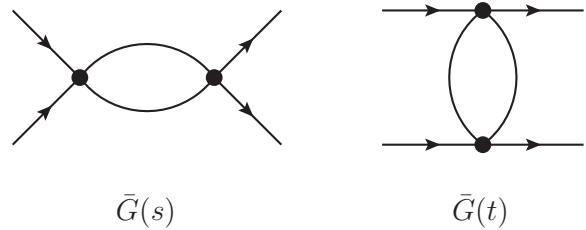}
%  \begin{picture}(60,80)(60,0)
%    \thicklines
%    \put(0,70){\line(1,-1){20}}
%    \put(0,30){\line(1,1){20}}
%    \put(40,50){\circle{40}}
%    \put(60,50){\line(1,1){20}}
%    \put(60,50){\line(1,-1){20}}
%
%    \put(100,70){\line(1,0){80}}
%    \put(100,30){\line(1,0){80}}
%    \put(140,50){\circle{40}}
%
%    \put(30,0){\text{$\bar G(s)$}}
%    \put(130,0){\text{$\bar G(t)$}}
%  \end{picture}
  \caption{Pion loops in the $s$ (left) and $t$ (right) channels contributing
    to the $O(p^4)$ chiral $\pi\pi$ scattering amplitude $A_4(s,t,u)$. 
    The $s$--channel loop
    gives rise to the right unitarity cut, whereas the $t$--channel
    (as well as $u$--channel) loop contributes to the left-hand cut in the partial wave.}
  \label{fig:loopdiagrams}
\end{figure}

Let us first 
note that the ${\cal O}(p^4)$ contribution to the 
$\pi\pi$ scattering amplitude $A_4(s,t,u)$, where the 
partial wave amplitude at the same order ($T_4^{IJ}(s)$) is obtained by projecting $A_4(s,t,u)$ on isospin $I$
and angular momentum $J$, has the 
form~\cite{Gasser:1983yg}
\begin{equation}
  \label{eq:A4}
  A_4(s,t,u)=B(s,t,u)+C(s,t,u)~,  
\end{equation}
where $C(s,t,u)$ is a second order polynomial
in $s$, $t$, and $u$ which contains the LECs. On the other hand, 
$B(s,t,u)$ contains the non--analyticities
coming from the one--loop diagrams with vertices from the $O(p^2)$
Lagrangian, 
\begin{equation}
  \begin{aligned}
    \label{eq:B}
    &B(s,t,u)=\frac1{6f_\pi^4}\Big[3(s^2-m_\pi^4)\bar G(s)\\
      &+\left\{t(t-u)-2m_\pi^2t+4m_\pi^2u-2m_\pi^4\right\}\bar G(t)\\
      &+\left\{u(u-t)-2m_\pi^2u+4m_\pi^2t-2m_\pi^4\right\}\bar G(u)\Big]~,
  \end{aligned}
\end{equation}
with
\begin{equation}
  \label{eq:Jbar}
  \bar G(s)=\frac1{16\pi^2}\left(\sigma(s)\log\frac{\sigma(s)+1}{\sigma(s)-1}-2\right)~,
\end{equation}
where we have defined $\bar G(s)\equiv G(s)-G(0)$.\footnote{This $\bar G$
relates to the $\bar J$ used in~\cite{Gasser:1983yg} as $\bar J=-\bar G$,
in accordance with our normalization, where the amplitudes have opposite
sign to those in~\cite{Gasser:1983yg}.} $G(s)$ is the loop function of two pions, Eq.~\eqref{loop}
with $s=P^2$. There are three loop
function contributions, proportional to $\bar G(s)$, $\bar G(t)$ and $\bar G(u)$,
each one coming from a pion loop in the $s$, $t$ and $u$--channels
respectively, as schematically shown in Fig.~\ref{fig:loopdiagrams}.
The $s$--channel loops are responsible for the right unitarity cut, and contain
the most important $L$ dependence of the amplitude. This $L$ dependence coming from
the unitarity cut is the one used by the L\"uscher/chiral unitary approach
method to obtain the phase-shift from the energy levels in a finite volume.
However, the $t$ and $u$--channel loops, which give rise to the left-hand
 cut when projecting into partial waves, give an extra dependence on $L$ (polarization corrections 
in the terminology of Ref.~\cite{Luscher:1990ux}) that
is neglected in the L\"uscher/chiral unitary approach method since it is
exponentially suppressed. 

Then, to obtain the IAM amplitudes in finite volume we have to 
replace $T_4(s)$ in Eq.~\eqref{eq:IAM} with $\widetilde T_4(s)$, which is obtained 
projecting into the corresponding partial wave the $\pi\pi$
scattering amplitude in finite volume $\widetilde A_4(s,t,u)$,
obtained from Eq.~\eqref{eq:A4} but replacing
the loop functions in Eq.~\eqref{eq:B} with their finite volume counterparts,
$\bar G_{FV}$. Following again the procedure in~\cite{alberto},
the finite volume loop functions are obtained from the infinite volume ones by
\begin{equation}
  \begin{aligned}
    \label{eq:Jtilde}
    \bar G_{FV}(z)=\bar G(z)+\lim_{p_\textrm{max}\to\infty}\Bigg[
      &\frac1{L^3}\sum_{p_i}^{p_\textrm{max}}I(p_i,z)\\
      &-\int\limits_{p<p_\textrm{max}}\frac{d^3p}{(2\pi)^3}I(p,z)\Bigg]~,
  \end{aligned}
\end{equation}
with $z=s,~t$ or $u$. Now, the energy levels in the box
are obtained from the poles in the scattering partial wave~\eqref{eq:IAM}, 
or equivalently, the zeros of $T_2(s)-\widetilde T_4(s)$.
From these energy levels at several values of $L$ one
can re-obtain the phase-shifts for the infinite volume with 
the L\"uscher/chiral unitary approach
method, and compare them with the exact infinite volume result to quantify the
effect of neglecting the $L$ dependence coming from the left-hand cut.

%%%%%%%%%%%%%%%%%%%%%%%%%%%%%%%%%%%%%%%%%%%%%%%%%%%%%%%%%%%%%%%%%%%%%%%%%%%%%%%%%%%%5
\subsection{The N/D method}

The case presented in subsection~\ref{sec:BS},
 can be put in the more general framework of the N/D
 method \cite{mandelstam,nsd,ollerulf,Albaladejo:2008qa}. 
 The amplitude was denoted by $T(s)$ in Eq.~\eqref{bse}. This master formula is obtained by solving algebraically the N/D method \cite{mandelstam,nsd,ollerulf,anpoller}, with the crossed cuts treated perturbatively, while the 
 right-hand cut is resummed exactly. The different chiral 
 orders of $V(s) = V_2(s) + V_4(s) + \ldots$ are calculated
  by matching $T(s)$ with the perturbative amplitudes $T_n(s)$.
   In this way, up to ${\cal O}(p^4)$,
\begin{align}\label{eq:chiralexpansion}
T(s) &= \frac{V(s)}{1 - V(s)G(s)} \nn\\
&=T_2(s) + T_4(s) +\ldots \nn\\
&= V_2(s) + V_4(s) + V_2(s)^2G(s)+\ldots~,
\end{align}
where the ellipsis indicates ${\cal O}(p^6)$ and higher orders in the expansion. It results then:
\begin{align}
V_2(s) & = T_2(s)~, \nn\\
V_4(s) & = T_4(s) - T_2(s)^2 G(s)~.
 \label{eq:v4}
\end{align}
The finite piece of the unitarity term in the $\pi\pi$ chiral amplitude is given by:
\begin{equation}
T_4^U(s) = T_2(s)^2 \bar{G}(s)~,
\end{equation}
with $\bar{G}(s)$ given in Eq.~\eqref{eq:Jbar}. In this way, the kernel $V(s) = V_2(s) + V_4(s)$ has no unitarity cut because:
\begin{equation}\label{eq:nocutinkernel}
T_4^U(s) - T_2(s)^2G(s) = T_2(s)^2(\bar{G}(s) - G(s))~,
\end{equation}
and the cut is cancelled in the r.h.s. of the previous equation. The full 
right-hand cut stems then from the denominator $1 - V(s)G(s)$ in Eq.~\eqref{bse}. 

In the infinite volume case, the LECs are fixed to the experiment, as
well as the subtraction constant $a$. We use here the central values of
the fit given in \cite{sigmaff}, for which the values of the finite and
scale independent LECs $\bar{l}_i$ are $\bar{l}_1 = 0.8 \pm 0.9$,
$\bar{l}_2 = 4.6 \pm 0.4$, $\bar{l}_3 = 2 \pm 4$, $\bar{l}_4 = 3.9 \pm
0.5$. In terms of the latter, the so-called renormalized LECs, which
depend on the renormalization scale, are $10^3 l_1^r = -2.8 \pm 0.9$,
$10^3 l_2^r = 2.5 \pm 0.8$, $10^3 l_3^r = 2 \pm 6$, 
$10^3 l_4^r = 3 \pm 3$, where the renormalization scale is chosen as $\mu = 770\
\text{MeV}$. The subtraction constant $a$ takes the value $a=-1.2 \pm
0.4$. We additionally note here that the same subtraction constant is
used for both channels, as required by isospin symmetry \cite{jido}.

In order to study the finite volume scattering, the same replacements as in the IAM and BS methods must be done. In particular, in the kernel $V(s) \to \widetilde{V}(s)$ no change is needed in $V_2(s)$, whereas $V_4(s)$ is changed to $\widetilde{V}_4(s)$,
\begin{equation}
\widetilde{V}_4(s) = \widetilde{T}_4(s) - T_2(s)^2 \widetilde{G}(s)~.
\end{equation}
Notice that, in view of Eq.~\eqref{eq:nocutinkernel}, there is no effect in the $s$-channel contributions to the kernel $\widetilde{V}(s)$. The volume dependence enters then in the kernel through the $t$- and $u$-channel loop functions, where the replacement $\bar{G}(z) \to \bar{G}_{FV}(z)$ in Eq.~\eqref{eq:Jtilde} for $z=t,u$ must be done. The $s$-channel volume-dependence enters then at the denominator of the amplitude $\widetilde{T}(s) = \widetilde{V}(s)/(1-\widetilde{V}(s)\widetilde{G}(s))$ through the function $\widetilde{G}(s)$, Eq.~\eqref{tonediff}, which gives the most important contribution to the aforementioned dependence, as in the case of the IAM method.

%%%%%%%%%%%%%%%%%%%%%%%%%%%%%%%%%%%%%%%%%%%%%%%%%%%%%%%%%%%%%%%%%%%%%%%%%%%%%%%%%%%%%%%%%%%%%%%%%%%
%%%%%%%%%%%%%%%%%%%%%%%%%%%%%%%%%%%%%%%%%%%%%%%%%%%%%%%%%%%%%%%%%%%%%%%%%%%%%%%%%%%%%%%%%%%%%%%%%%%%%%%
\section{Results}
\label{sec3}

As already explained, the main aim of the present work is to quantify
the effect of the dependence of the different potentials considered
on the size of the box, $L$. Hence, we are going to compare the $L$
dependence 
of the N/D and
the IAM method with that of the BS, which kernel does not depend on $L$.

\begin{figure}[!h]
\begin{center}
\includegraphics[width=0.95\linewidth]{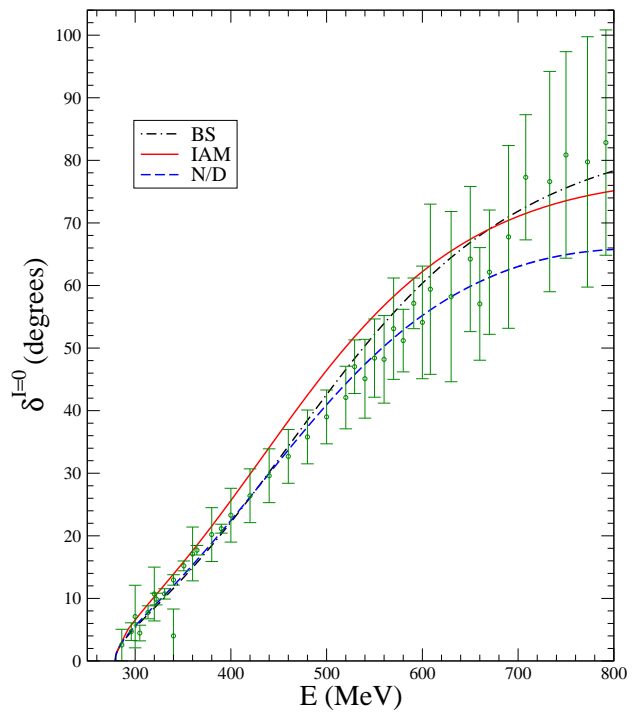}
\caption{Isospin I=0, s-wave, $\pi\pi\to\pi\pi$ phase-shifts for the 
three
different models considered: solid, dashed and dot-dashed lines correspond to 
 IAM, N/D and BS, respectively. The experimental data are from
  Refs.~\cite{hyams,kaminski,grayer,bnl,na48}.}
\label{fig:pasheshifts_I0_inf_vol}
\end{center}
\end{figure}

First we show in Fig.~\ref{fig:pasheshifts_I0_inf_vol} the results for
the $\pi\pi$ phase-shifts in s-wave and $I=0$ for the three different
models in infinite volume. The IAM and N/D  results (solid and dashed lines, respectively) are the fits
explained in the previous section and the BS (dot-dashed line) is fitted in this 
work to the experimental data \cite{hyams,kaminski,grayer,bnl,na48} shown in the figure up 
to $800\mev$. 
The IAM and N/D approaches are essentially equivalent at low energies
but differ slightly as the energy increases.
Thus the difference between the IAM and N/D phase shifts 
 is mainly due to the
different set of data used in the fit and it also gives an idea of
the theoretical uncertainty. The BS approach produces a curve in between the other two, closer to 
the N/D at low energies and to the IAM at higher energies. In any case, the different models are
compatible within the experimental uncertainties.
 Let us note that what matters for the discussions in the present work is not the
actual values of the phase-shifts at infinite volume but the relative
change when going to the finite box.

\begin{figure}[!h]
\begin{center}
\includegraphics[width=0.95\linewidth]{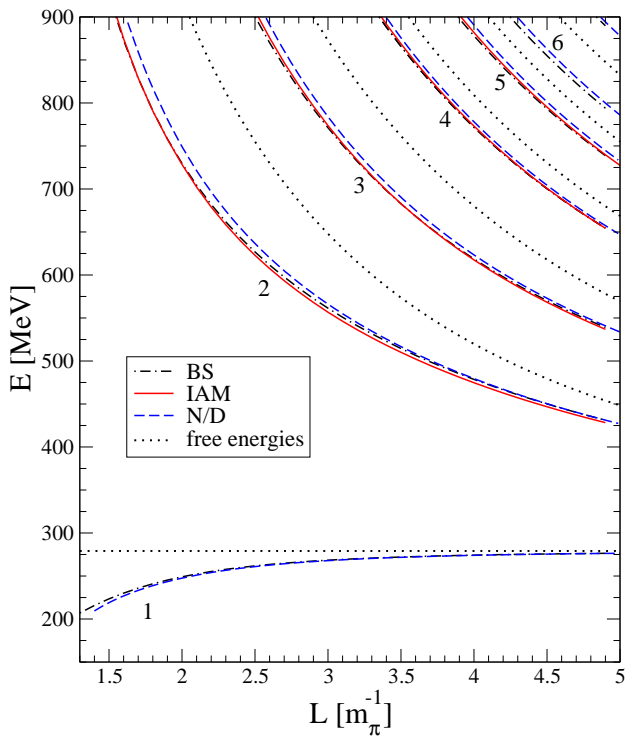}
\caption{The first energy levels as a function of the cubic box size 
$L$ for the three different models considered for $I=0$.
The dotted lines indicate the free $\pi\pi$ energies in the box. The rest of the lines 
correspond to IAM, N/D and BS as in Fig.~\ref{fig:pasheshifts_I0_inf_vol}.
}
\label{fig:EvsLI0}
\end{center}
\end{figure}

In Fig.~\ref{fig:EvsLI0} we show the energy levels for different
values of the cubic box size, $L$,
for the different models which have been obtained from 
the zeroes of the scattering
amplitudes in the finite box as explained in the previous section.
 The
dotted lines represent the free $\pi\pi$ energies in the box, while the others lines 
correspond to IAM, N/D and BS as in Fig.~\ref{fig:pasheshifts_I0_inf_vol}.

The differences are very small for the largest values of $L$ shown in
the plot but are  more important for smaller values of $L$,
specially between the N/D and IAM methods. The BS approach produces a curve in between 
the other two, closer to the N/D. The IAM and BS are more
similar for larger values of energies as can also be seen in the phase
shifts,
Fig.~\ref{fig:pasheshifts_I0_inf_vol}.
As an example of small $L$, we note that
 for $L=1.7m_\pi^{-1}$ the difference between N/D and IAM
is about $30\mev$.

An actual lattice calculation would provide some points over
analogous trajectories in the $E$~vs.~$L$ plots.
The ``inverse problem'' is the problem of getting the actual
scattering amplitudes (and hence by-product magnitudes like
phase-shifts) in the infinite space from data produced by lattice QCD 
 consisting of points in plots of  $E$~vs.~$L$ over
the energy levels in the box.
For points in these levels the amplitude in the infinite volume 
can be extracted from the generalization
of the L\"uscher formula, as explained in the previous sections,
   \be
T(E)=\frac{1}{\widetilde G(E)-G(E)} \ . 
\label{Lusch}
\ee 

\begin{figure}[!h]
\begin{center}
\includegraphics[width=0.95\linewidth]{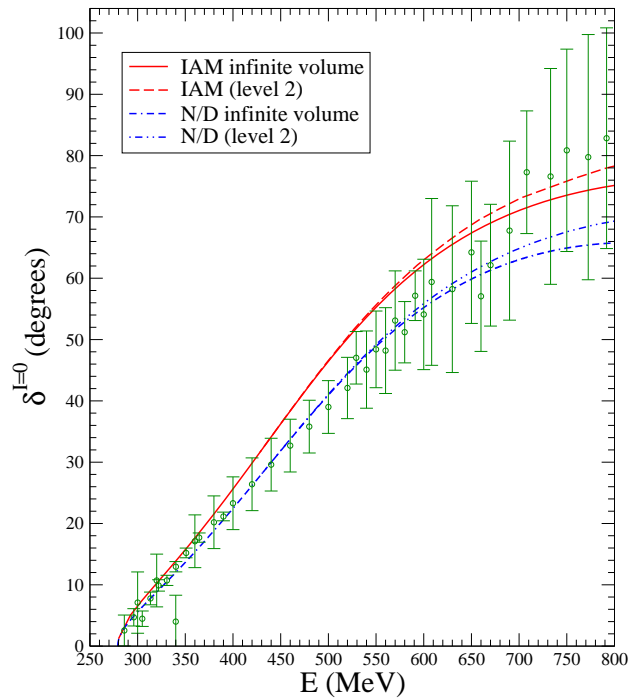}
\caption{Solution of the inverse problem for $I=0$ for the IAM and
N/D methods. The BS result is the same as in the infinite volume case
and thus we do not show it in the figure. We show the results
obtained only from level 2   of Fig.~\ref{fig:EvsLI0} since the results
with levels $>2$ almost overlap with the infinite volume line. For the meaning of each line consult 
the inset in the figure. }
\label{fig:fig_def_Luscher}
\end{center}
\end{figure}
In Fig.~\ref{fig:fig_def_Luscher} 
we show the phase-shifts obtained for the different methods
implementing the ``inverse problem'' analysis (or ``reconstructed'' results)
 with  Eq.~(\ref{Lusch})
and from the $E$~vs.~$L$ plot. 
For the $BS$ model
the results are independent of the level used for a given $E$, since the
potential does not depend on $L$, and they are equal to the 
infinite volume result. Therefore we do not show 
the BS result since it is the same as
 in Fig.~\ref{fig:pasheshifts_I0_inf_vol}.
For the IAM and N/D methods
the results depend on the level chosen for a given $E$ since the
potentials depend on $L$ as explained in the
previous sections. Actually, for levels
$>2$ of Fig.~\ref{fig:EvsLI0} the results are almost
 equal to the infinite volume results and
hence we do not show them in the figure since they would almost 
overlap with
the infinite volume line.
This is because,
as seen  in Fig.~\ref{fig:EvsLI0}, for the higher energies these levels
imply large values of $L$.
Indeed, for energies below $800\mev$
this implies values of $L$ higher than about $3m_\pi^{-1}$. For the
results obtained with level 2, the phase-shifts differ in about
5\% of the result in the infinite volume at the higher energies
considered.
For $E\sim 800\mev$ this implies $L$ values slightly smaller than
$2m_\pi^{-1}$, as can be seen in Fig.~\ref{fig:EvsLI0}.
 It is worth
noting that the effect of the dependence on $L$ of the models with
left-hand cut go in the same direction and are of similar size in spite
of the different models used. This  gives us confidence that
the actual $L$ dependence of the left-hand cut is properly considered
and the real effect of
any realistic model would be of the order obtained
in the present work.
An analysis  with 
Eq.~(\ref{Lusch}) applied to actual lattice results of $E$ versus $L$
levels would neglect the possible $L$ dependence of
the potential and hence the errors from the $L$ dependence of the
left-hand cut
would be of the order of the differences shown in
the figure. Note also that the $L$ dependence of the results 
are smaller than the initial difference between the N/D and IAM
 themselves
and also lower than the experimental uncertainties. Therefore, an
actual lattice calculation should care about this $L$ dependence
only if it aims at getting errors smaller than
the effect obtained in the present work.

\begin{figure}[!h]
\begin{center}
\includegraphics[width=0.95\linewidth]{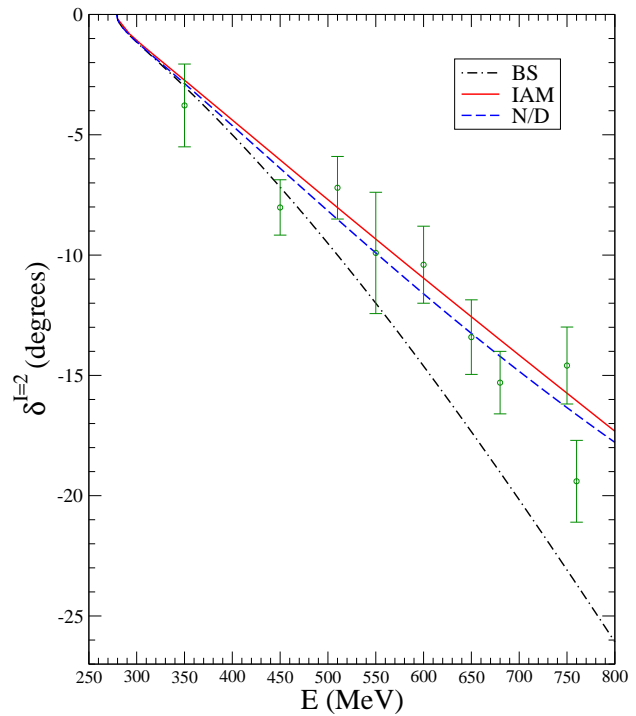}
\caption{Isospin I=2, s-wave, $\pi\pi\to\pi\pi$ phase-shifts for the 
three
different models considered. The experimental data are from 
Refs.~\cite{Losty:1973et,Hoogland:1977kt}. See the inset in the figure for the 
correspondence between the different lines and the approach used.}
\label{fig:pasheshifts_I2_inf_vol}
\end{center}
\end{figure}

\begin{figure}[!h]
\begin{center}
\includegraphics[width=0.95\linewidth]{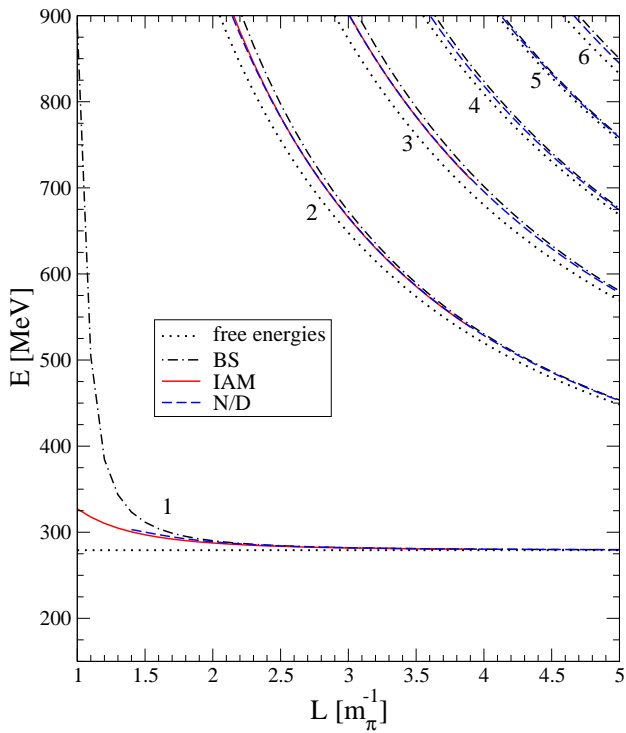}
\caption{The first energy levels as a function of the cubic box size 
$L$ for the three different models considered for $I=2$.
The meaning of the different lines is as in Fig.~\ref{fig:EvsLI0}. } 
\label{fig:EvsLI2}
\end{center}
\end{figure}

\begin{figure}[!h]
\begin{center}
\includegraphics[width=0.95\linewidth]{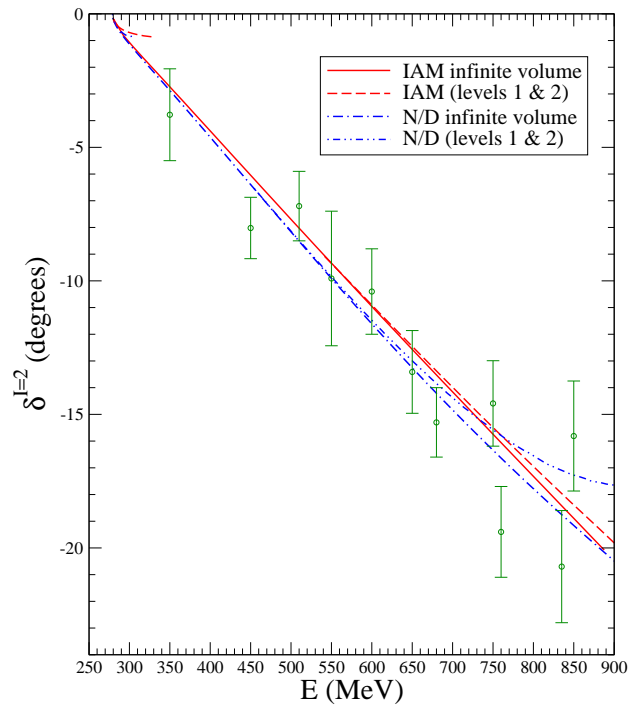}
\caption{Solution of the inverse scattering problem with $I=2$ for the IAM and
N/D methods. The BS result is the same as in the infinite volume case
and thus we do not show it in the figure. We show the results
obtained only from level 1 and 2   of Fig.~\ref{fig:EvsLI2}
since the results
with levels $>2$ almost overlap with the infinite volume line. For the meaning of the 
lines consult the inset in the figure.}
\label{fig:fig_I2_def_Luscher}
\end{center}
\end{figure}

In Figs.~\ref{fig:pasheshifts_I2_inf_vol}, \ref{fig:EvsLI2} and
\ref{fig:fig_I2_def_Luscher} we show for the $I=2$ case the same results as in 
Figs.~\ref{fig:pasheshifts_I0_inf_vol} to \ref{fig:fig_def_Luscher} for  $I=0$. 
In Fig.~\ref{fig:pasheshifts_I2_inf_vol} we see that the
IAM and N/D methods provide very similar results and compatible with
the experimental data while the BS approach
 gets worse phase-shifts.
 This is because in the IAM
and N/D the left-hand cut is included perturbatively order by order,
unlike the BS model, and in this channel the left-hand cut is more
relevant. 
In Fig.~\ref{fig:EvsLI2} we show the energy levels in the box for this
channel.

Now both IAM and N/D provide similar results.
In Fig.~\ref{fig:fig_I2_def_Luscher} we show the solution
of the inverse problem for the phase-shifts.
We see that the N/D method provides a higher $L$ dependence for large
values of the energies, unlike IAM.
 At $800\mev$ the difference is about $10\%$ for the N/D and $2\%$
for the IAM. The difference in the phase-shifts between the two
approaches is
large in spite of the energy levels being very similar. This is 
because the
energy levels are very close to the free case, unlike the $I=0$ case, and
then the $\widetilde G$ function is very steep. This makes that
 small variations in $E$
provide large variations in $\widetilde G$.

In usual inverse problem analysis from actual lattice results, it is common
to use the L\"uscher formula
\cite{luscher,Luscher:1990ux} which,
as explained in section~\ref{sec:BS}, is an approximation 
to that used in the present work, Eq.~(\ref{extracted_1_channel}).
Therefore it is worth studying what is the error made in the 
reconstructed phase-shifts if one uses the L\"uscher equation instead of
the exact one. In Ref.~\cite{misha} it was shown that the L\"uscher
method can be reproduced if in Eq.~(\ref{prop_contado}) one substitutes 
\be
I(p,s)=
\frac{1}{\omega(\vec p)}\,\,
\frac{1}
{s-4\omega(\vec p)^2}.
\label{prop_contado2}
\ee
by
\be
I(p,s)=
\frac{1}{2\sqrt{s}}\,\,
\frac{1}
{p_{\textrm{ON}}^2-\vec p\,^2}.
\label{eq:prop_Luschpure}
\ee
where $p_{\textrm{ON}}=\frac{E}{2}\sqrt{1-\frac{4m^2}{s}}$.
\begin{figure}[!h]
\begin{center}
\includegraphics[width=0.95\linewidth]{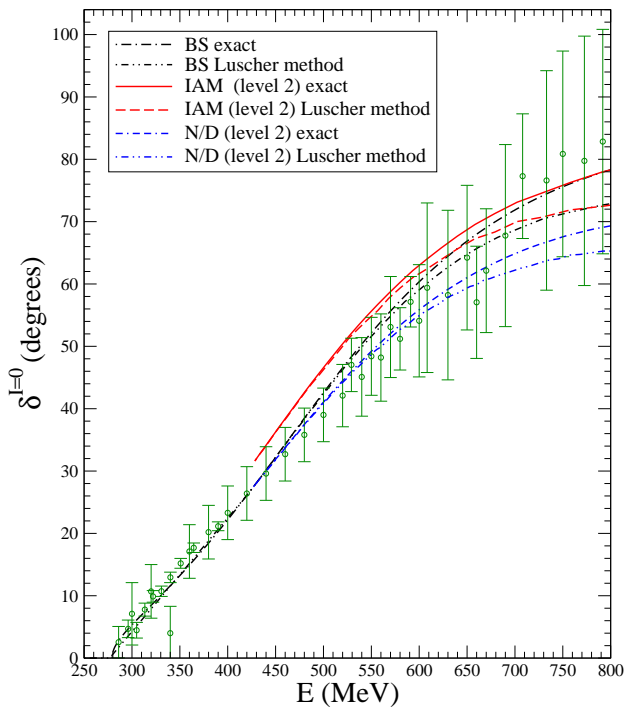}
\caption{Difference between the exact inverse method formula,
Eq.~(\ref{extracted_1_channel}), and the approximated 
L\"uscher formula. The approach corresponding to each line is given in the inset in the figure.}
\label{fig:fig_def_Luscher_puro}
\end{center}
\end{figure}

In Fig.~\ref{fig:fig_def_Luscher_puro} we
 show the effect in the isospin 0 phase-shifts
of using the pure L\"uscher
method, Eq.~(\ref{eq:prop_Luschpure}), instead of the exact one,
Eq.~(\ref{prop_contado2}).
(For the isospin 2 case the effect is small and thus we do not show any
plot.)
The difference is significant only for phase-shifts extracted from 
level 2 of 
Fig.~\ref{fig:EvsLI0} since the difference is only relevant for small
values of $L$. Therefore we only plot results extracted from level 2.
The difference between the exact method and the L\"uscher one is similar
for all the three different models for the potential.
The size of the difference is similar to the one from the $L$ dependence
of the potential discussed above but goes in the opposite direction.
Therefore they tend to compensate each other by chance.

%%%%%%%%%%%%%%%%%%%%%%%%%%%%%%%%%%%%%%%%%%%%%%%%%%%%%%%%%%%%%%%%%%%%%%%%%%%%
\section{Summary}
\label{sec4}

In this paper we have faced for the first time in the literature the problem of the
presence of the left-hand cut of the $\pi\pi$ amplitude for 
the evaluation of phase-shifts from lattice QCD results using L\"uscher's
approach. The $t-$ and $u-$channel terms can be taken into account in a
field theoretical approach by means of the IAM, or N/D NLO methods,
leading to good reproductions of the scattering data. Results from
lattice QCD should contain all the dynamics and, as a consequence,
should account for these effects too. However, the method to go from the
discrete energy level in a box from lattice simulations to the phase
shifts for scattering in the infinite volume case requires the use of
L\"uscher's approach, or its improved version of \cite{misha}, both of
which rely upon the existence of a volume independent potential.  Yet,
the terms contributing to the left-hand cut, containing loops in the
$t-$
and $u-$channels, are explicitly volume dependent. In this work we have
investigated the errors induced by making use of \cite{luscher} or
\cite{misha} in the reproduction of phase-shifts from the energy spectra
of lattice calculations in the finite box. We have found that in the
case of $\pi \pi $ scattering in s-wave, both for $I=0$ and $I=2$, the
effect of the $L$ dependence in the potential is smaller than the
typical errors from the experimental phase-shifts or the differences
between the three models that we have used, the IAM, N/D NLO and
BS LO. This is good news for lattice calculations since one of the
warnings not to go to small values of $L$ was the possible $L$ dependence of the potential
which in some cases, like in the present one, we know that exists. We
found that it is quite safe to ignore this dependence for
 $L> 2.5m_{\pi}^{-1}$, and even with values of $L$ around $1.5 -2 m_{\pi}^{-1}$
the errors induced are of the order of 5\%. 

On the other hand we have quantified the error made
by using the pure L\"uscher formula instead
of the exact one, Eq.~(\ref{Lusch}). The effect in the phase-shifts of
this approximation tends to compensate, by chance, the effect of neglecting
the  $L$
dependence in the potential discussed so far.

All these findings, together with
the use of the approach of \cite{misha} that also eliminates $L$ depended
terms (exponentially suppressed) from the L\"uscher's approach, can
encourage the performance of lattice calculations with smaller size
boxes with the consequent economy in the computing time.

\section*{Acknowledgments}
 This work is partly supported by DGICYT contracts  FIS2006-03438,
 the Generalitat Valenciana in the program Prometeo 2009/09, MEC  FPA2010-17806, the Fundaci\'on S\'eneca  11871/PI/090 and 
the EU Integrated Infrastructure Initiative Hadron Physics
Project under Grant Agreement n.227431.

\end{document}